\documentclass[reprint,amsmath,amssymb,aps,]{revtex4-1}

\usepackage{graphicx}% Include figure files
\usepackage{dcolumn}% Align table columns on decimal point
\usepackage{bm}% bold math

\begin{document}

\title{An order-disorder phase transition in black-hole star clusters}

\author{Jihad Touma}
\affiliation{Department of Physics, American University of Beirut, PO Box 11-0236, Riad El-Solh, Beirut 11097 2020, Lebanon}
\author{Scott Tremaine}
\email{tremaine@ias.edu}
\affiliation{Institute for Advanced Study, 1 Einstein Drive, Princeton NJ 08540}
\author{Mher Kazandjian}
\affiliation{Leiden Observatory, Leiden University, PO Box 9513, 2300 RA Leiden, The Netherlands}

\newcommand{\bfr}{\boldsymbol{r}}
\newcommand{\pc}{\mbox{\,pc}}
\newcommand{\yr}{\mbox{\,yr}}
\newcommand{\Gyr}{\mbox{\,Gyr}}
\newcommand{\msun}{\,M_\odot}
\newcommand{\kms}{\mbox{\,km s}^{-1}}
\newcommand{\erms}{e_\mathrm{rms}}
\newcommand{\ermss}{e_\mathrm{rms}^\mathrm{sph}}
\newcommand{\ecrit}{e_\mathrm{crit}}

% Include the date command, but leave its argument blank.

\date{\today}

\begin{abstract}
  The centers of most galaxies contain massive black holes surrounded by dense star clusters. The structure of these clusters determines the rate and properties of observable transient events, such as flares from tidally disrupted stars and gravitational-wave signals from stars spiraling into the black hole. Most estimates of these rates enforce spherical symmetry on the cluster. Here we show that, in the course of generic evolutionary processes, a star cluster surrounding a black hole can undergo a robust phase transition from a spherical thermal equilibrium to a lopsided equilibrium, in which most stars are on high-eccentricity orbits with aligned orientations. The rate of transient events is expected to be much higher in the ordered phase. Better models of cluster formation and evolution are needed to determine whether clusters should be found in the ordered or disordered phase.

  \vspace{0.3cm}\texttt{https://journals.aps.org/prl/accepted/7d07fY20P5414669399484c5b2c3677a256e33d4a}
\end{abstract}

%% \pacs{Valid PACS appear here}% PACS, the Physics and Astronomy
                             % Classification Scheme.

\maketitle

\section{Introduction}

The central black-hole mass in galaxies, $M_\bullet$, is tightly correlated with the large-scale properties of the galaxy. In particular the mean-square line-of-sight velocity of the stars, $\sigma^2$, is related to the black-hole mass through \cite{kh13} $M_\bullet\simeq 3\times 10^8M_\odot (\sigma/200\kms)^{4.4}$ over the range $10^6\lesssim M_\bullet/\msun \lesssim 10^{10}$. The orbits of stars are dominated by the gravitational field of the black hole within its sphere of influence, of radius $r_\mathrm{infl}=GM_\bullet/\sigma^2\simeq 20\pc\,(M_\bullet/10^8M_\odot)^{0.55}$. For comparison, the event horizon of the black hole is at least five orders of magnitude smaller,  $r_\bullet=2GM_\bullet/c^2=9.6\times10^{-6}\pc\, (M_\bullet/10^8\msun)$ where $c$ is the speed of light. We may refer to the stars in the region between $r_\bullet$ and $r_\mathrm{infl}$ as the black-hole star cluster.

Within this sphere stars travel on nearly Keplerian orbits with period $2\pi t_\mathrm{dyn}(a)$, where $a$ is the semimajor axis and the dynamical time $t_\mathrm{dyn}(a)=1.5\times 10^3\yr(a/\pc)^{3/2}(10^8\msun/M_\bullet)^{1/2}$. Over the age of the galaxy, $\sim 10\Gyr$, gravitational forces between individual stars gradually randomize their orbits. The most important randomization process in black-hole clusters is resonant relaxation \cite{rt96,st16,tal17,bof18}, which results from the time-averaged force exerted by each stellar orbit on the other stars. Resonant relaxation is faster than relaxation due to close encounters between stars, also called two-body relaxation, by a factor $\simeq 0.15 M_\bullet/[M_\star(r) g(r)\log\Lambda]$. Here $M_\star(r)$ is the mass of stars within radius $r$, $\log\Lambda\simeq \log M_\bullet/m$ is the Coulomb logarithm, $m$ is a typical stellar mass, and $g(r)\simeq \max[1,4(r_\bullet/r) M_\bullet/M_\star(r)]$ accounts for the suppression of resonant relaxation by relativistic apsidal precession (the numerical factors are for a typical eccentricity of 0.7). For example, in the well-studied star cluster surrounding the $4\times10^6\msun$ black hole at the center of the Milky Way, resonant relaxation is the dominant relaxation mechanism---faster than two-body relaxation---over a wide range of radii, from about 0.001$\pc$ to 0.1$\pc$. Resonant relaxation is also relatively fast: the minimum resonant-relaxation time, at semimajor axes near $0.01\pc$, is only $10^8\mbox{\,yr}$ or 1 per cent of the age of the galaxy \cite{bof18,kt11,geg10}.

The distribution of old stars in the Milky Way's black-hole star cluster is approximately spherical. Spherical models are almost always taken for granted in theoretical studies of black-hole star cluster dynamics \cite{bw76,ck78}.

Because resonant relaxation results from orbit-averaged forces, the Keplerian energies or semimajor axes of the stellar orbits are conserved \cite{rt96}. Therefore resonant relaxation leads to an equilibrium phase-space distribution that maximizes the entropy, subject to the constraint that the semimajor axes are conserved \cite{footnote0}. Additional conserved quantities are the total mass and angular momentum as well as the total non-Keplerian energy, which arises from the self-gravity of the stars and relativistic corrections to the Kepler Hamiltonian.  The goal of this paper is to investigate these maximum-entropy states, which black-hole star clusters should occupy if their age is longer than the resonant-relaxation time. On much longer timescales, the semimajor axis distribution of the cluster evolves due to two-body relaxation, but the other orbital elements should still be in the maximum-entropy state consistent with the slowly evolving semimajor axis distribution. We shall find that these states exhibit a robust phase transition, from a disordered spherical state to an ordered lopsided state, as the self-gravitational energy of the stars is lowered. 

\section{Numerical models}

In all of the models described here, the total angular momentum of the cluster was zero and relativistic precession was neglected (see \S\ref{sec:discussion} for a discussion of relativistic effects). The self-gravitational energy of the stars was computed by evaluating the orbit-averaged potential energy between pairs of Kepler ellipses, using either a spherical-harmonic expansion (\S\S \ref{sec:one} and \ref{sec:four}) or Gauss's method (\S\S \ref{sec:two} and \ref{sec:three}). This approach should be valid as long as the precession time $t_\mathrm{prec}\sim t_\mathrm{dyn}M_\bullet/M_\star$ (the time needed for the line of apsides to precess one radian due to the mean gravitational field of the cluster) is much longer than the orbital time, which requires that the total mass in stars is much less than the black-hole mass, $M_\star(r)\ll M_\bullet$. 

\subsection{Maximum-entropy states of a mono-energetic cluster}

\label{sec:one}

First, we found the maximum-entropy states of a cluster in which all the stars have a single semimajor axis $a_0$. We call this a ``mono-energetic'' cluster since the Keplerian energy is $-GM/(2a_0)$. This assumption is clearly unrealistic, but provides a simple test case that can be explored thoroughly \cite{tt14} and in which the evolution is dominated by orbit-averaged forces even though all stars have the same period. We also assume that the cluster is axisymmetric, so the orbit-averaged distribution function may be determined on a three-dimensional grid in phase space (eccentricity, inclination, and argument of periapsis). The corresponding gravitational potential is obtained using a spherical-harmonic expansion up  to  order $l_\mathrm{max}=8$. Terminating the expansion at finite $l_\mathrm{max}$ softens the gravitational force, but we have confirmed that the results below are insensitive to the precise value of  $l_\mathrm{max}$. We ignore any loss of stars through tidal disruption or consumption by the central black hole. 

These systems constitute a family of microcanonical ensembles that can be characterized by a single control parameter $E$, the self-gravitational energy of the stellar system in units of $GM_\star^2/a_0$. Spherically symmetric systems---those in which the distribution function depends only on the eccentricity $e$---provide an important reference point. Since the distribution function depends on only one variable and the gravitational potential is spherical it is straightforward to calculate the maximum-entropy spherical distribution at a given mass and energy. We find that the rms eccentricity $\ermss$ varies monotonically with energy, from zero when $E=-1/2$ (all stars on circular orbits) to unity when $E=-0.29736$ (all stars on radial orbits). Physically, when $\ermss$ is small, the stellar mass distribution has less radial width so the self-gravitational energy is more negative. Thus $\ermss$ can be used as an alternative control parameter. Another alternative is the inverse temperature $\beta$, defined for the microcanonical ensemble such that the distribution function is $\propto \exp(-\beta H)$ where $H$ is the Hamiltonian. For spherical systems, $\beta$ declines monotonically with increasing $E$ and is zero for $E=-0.3559$, corresponding to an ergodic distribution function that is independent of all orbital elements other than the semimajor axis and has $\ermss=2^{-1/2}=0.707$.

\begin{figure}[ht!]
\centering
\includegraphics[width=0.48\textwidth]{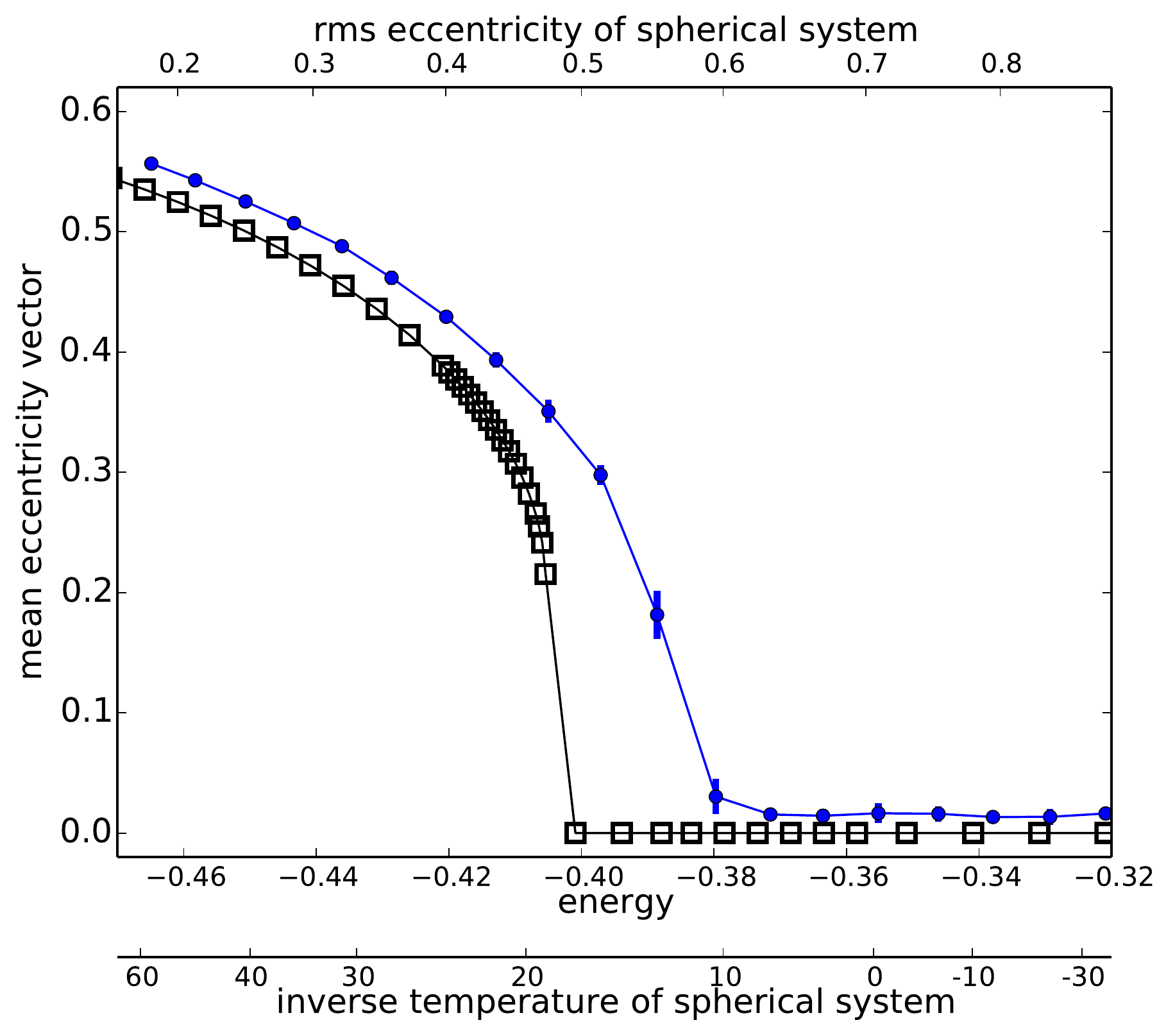}
\caption{Lopsided phase transition in a maximum-entropy cluster of stars. The cluster stars have a single semimajor axis and zero total angular momentum. The vertical axis shows the magnitude of the mean eccentricity vector. The horizontal axes show the self-gravitational energy, as well as the rms eccentricity $\ermss$ and inverse temperature $\beta$ of the spherical system with the same energy. Open black squares show the axisymmetric system described in \S\ref{sec:one}, while filled blue circles with error bars are from the MCMC simulation of a cluster of 2048 stars described in \S\ref{sec:two}. The units of $E$ and $\beta^{-1}$ are $GM_\star^2/a_0$.
\label{fig:one}}
\end{figure}

The eccentricity or Runge--Lenz vector of an orbit points from the central mass towards periapsis and has magnitude equal to the (scalar) eccentricity $e$. The mean eccentricity vector $\langle\mathbf{e}\rangle$ of the cluster is zero when the system is spherical and serves as an order parameter. 

We represent a mono-energetic cluster by the values of the distribution function on a $(16)^3$ grid in phase space, and choose these values to maximize the entropy subject to the constraints that the distribution function is non-negative, the total angular momentum is zero, and the self-gravitational energy $E$ is fixed. The open squares in Figure \ref{fig:one} show the mean eccentricity vector of the maximum-entropy state as a function of $E$, $\beta$ (bottom axes), and $\ermss$ (top axis). For $E\gtrsim -0.40$, $\beta\lesssim 18$, or $\ermss\gtrsim 0.48$, $\langle\mathbf{e}\rangle=0$ and the system is spherical. For lower energies or rms eccentricities, or larger inverse temperature, the maximum-entropy system is lopsided and $|\langle\mathbf{e}\rangle|$ is non-zero. In other words there is an order-disorder phase transition at $\beta\simeq 18$. Just above the critical inverse temperature (i.e., to the left of the transition in Figure \ref{fig:one}), spherical systems are metastable, that is, they are located at a local but not global maximum of the entropy. Linear stability analysis shows that spherical systems lose their metastability at $E=-0.44$, through an unstable $l=2$ mode. As the system is cooled below the critical temperature (i.e., moves to the left of $E=-0.40$ in Figure \ref{fig:one}), the inverse temperature $\beta$ grows, the orbits in the ordered state become more eccentric and their lines of apsides become more closely aligned, so the mean eccentricity vector grows. These results apply to axisymmetric clusters, in which the lopsided equilibria break the mirror symmetry around the equatorial plane; we have also experimented with non-axisymmetric clusters but did not find any novel behavior \cite{footnote2}. 

\subsection{Markov chain Monte Carlo simulations of a mono-energetic cluster}

\label{sec:two}

We used Markov chain Monte Carlo (MCMC) simulations to construct microcanonical equilibria of a mono-energetic cluster with a given value of $E$, the energy of the stars due to their self-gravity. This approach relies on a Markov chain of pairwise interactions to guide the system to a stationary maximum-entropy state, and is distinct from (and significantly cheaper than) the dynamical simulations in \S\ref{sec:three}. The simulations used $N=2048$ stars, and the orbit-averaged gravitational interaction between each pair of stars was computed using Gauss's method \cite{ttk09} with a softening length $b=0.1a_0$. At each step in the MCMC simulation we chose a random orbit pair and changed the orbital elements of the pair randomly subject to the constraint that the total angular momentum is conserved. We employed Creutz's demon algorithm \cite{cre83} to constrain $E$ at a fixed value.  Typically $\sim 10^6$ steps were needed to reach an equilibrium state, and the equilibrium properties were sampled over another $\sim 10^6$ steps after equilibrium was reached. The simulations, shown as the solid circles with error bars in Figure \ref{fig:one}, exhibit the same order-disorder transition the model in \S \ref{sec:one}. The only significant difference is a small shift in the location of the phase transition---$E=-0.38$ versus $E=-0.40$---which probably
is due to a combination of finite-size effects and gravitational softening in the MCMC simulations.

\begin{figure}[ht!]
\centering
\includegraphics[width=0.48\textwidth]{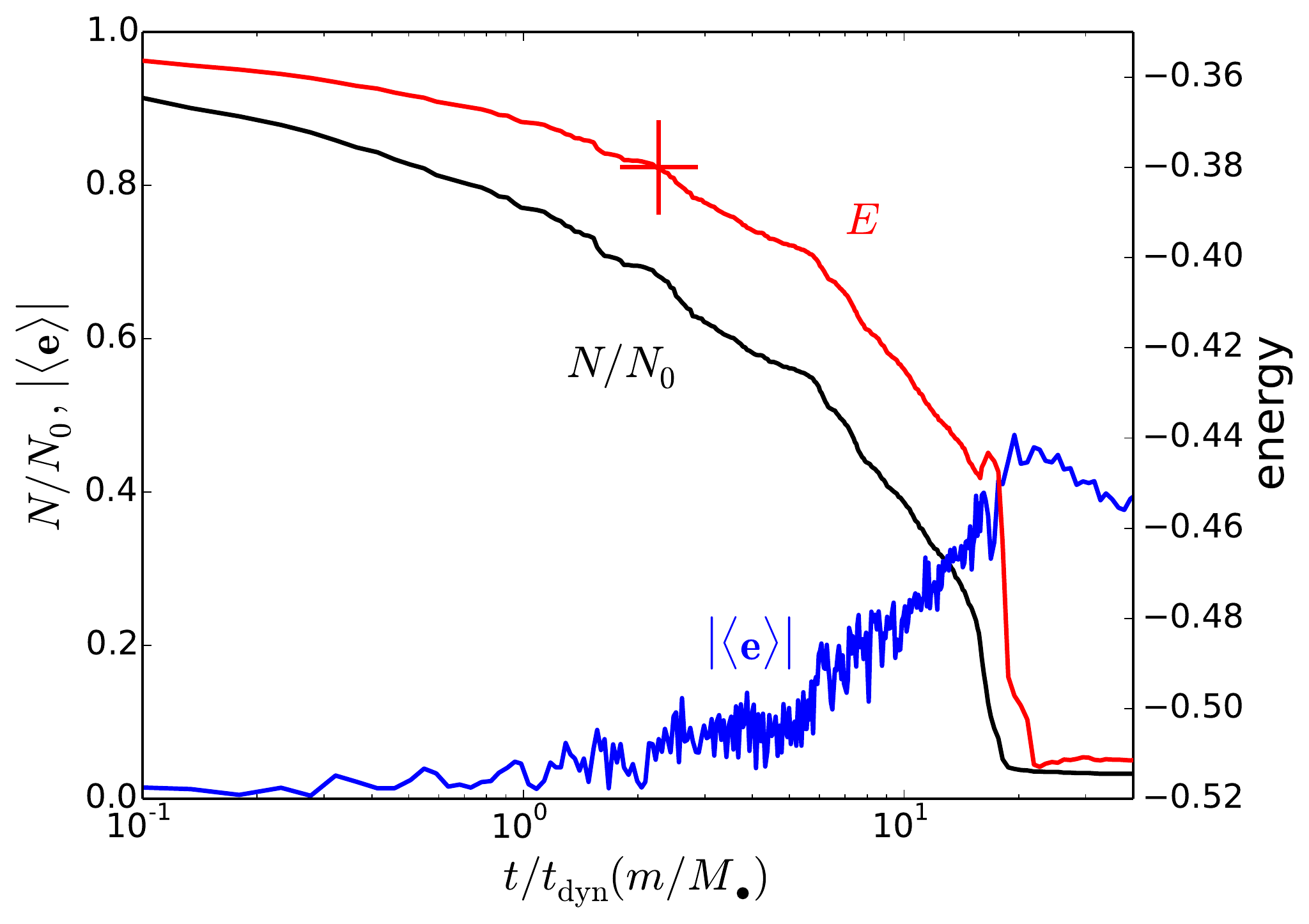}
\caption{Secular evolution of a cluster of stars. The initial cluster is composed of $N_0=2048$ stars with a single semimajor axis $a_0$ and zero total angular momentum. The evolution is followed with an $N$-wire code \cite{ttk09} as described in \S\ref{sec:three}. Stars passing within $0.01a_0$ of the central black hole are removed, so the number $N$ of stars declines with time (black curve). The energy $E$ in units of $GM_\star^2(t)/a_0$ also declines with time (red curve), and the mean eccentricity vector $|\langle\mathbf{e}\rangle|$ grows (blue curve). The bottom axis shows the time in units of $t_\mathrm{dyn}M_\bullet/m$, the dynamical time multiplied by the ratio of black-hole to stellar mass, which apart from dimensionless constants equals the resonant relaxation time when relativistic precession is ignored. The red cross marks the critical energy for the phase transition, $E=-0.38$, according to the MCMC simulation of Figure \ref{fig:one}. \label{fig:two}}
\end{figure}

\subsection{$N$-wire evolution of a mono-energetic cluster} 

\label{sec:three}

We followed the secular evolution of a mono-energetic cluster of $N=2048$ stars, again using Gauss's method with softening to evaluate the forces between stars and the resulting evolution of their orbits \cite{ttk09}.  The stars initially had an ergodic ($\beta=0$) distribution in the canonical phase-space coordinates other than semimajor axis. In addition, to mimic the loss of stars to the black hole, we removed any star that passed through periapsis at a distance $<0.01a_0$ and instantaneously added its mass to the central black hole. With this prescription, there is a steady decay in the number $N$ and mass $M_\star$ of the stars; in parallel, there is a steady decline in the self-gravitational energy $E$ (in units of $GM_\star^2/a_0$) as high-eccentricity orbits, which have higher energies, are preferentially lost to the black hole. The results are shown in Figure \ref{fig:two}: as the energy declines below the critical energy $E=-0.38$ found in calculation 2 (the red cross), the amplitude of the mean eccentricity vector starts to grow, achieving a maximum of $\sim 0.4$. The loss region in this simulation is unrealistically large for semimajor axes that are typical of the regions of black-hole clusters in which relativistic precession can be neglected. We have conducted experiments with smaller $N$ that have smaller loss regions, and these exhibit a similar lopsided transition but over a longer timescale. We conclude that the loss of high-eccentricity, high-energy stars cools the cluster so $\beta$ grows, leading to a transition to an ordered state. 

Analogous dynamical simulations with a Bahcall--Wolf distribution of semimajor axes \cite{bw76}, $dN\sim a^{1/4}da$, also exhibit the transition to a lopsided state. 

\subsection{Maximum-entropy states of a self-similar cluster} 

\label{sec:four}

Finally, we investigated the canonical equilibria of a self-similar cluster. By choosing the distribution of semimajor axes to be $dN(a) \propto da/a$ and the mass of a star to vary with semimajor axis as $m(a) \propto a^{1/2}$, we may ensure that the canonical equilibrium at each semimajor axis is identical.  Once again we find that as a spherical cluster is cooled it undergoes a phase transition to an ordered state in which the stellar orbits have high eccentricity and nearly aligned apsides. The transition occurs at $\ermss\simeq 0.6$. 

\section{Discussion}
\label{sec:discussion}

These phase transitions are distinct from dynamical instabilities. Because the self-gravity of the stars is weaker than the gravitational force from the black hole, the only possible dynamical instabilities in this region occur over the precession time $t_\mathrm{prec}\sim t_\mathrm{dyn}M_\bullet/M_\star \gg t_\mathrm{dyn}$. The cluster is stable on this timescale so long as the phase-space distribution of the stars is monotonic in angular momentum and relativistic effects are negligible \cite{tre05}; it can be unstable if the distribution function is non-monotonic, but maximum-entropy states are always monotonic in the absence of relativistic effects \cite{footnote1}.

The simple model systems in \S\S\ref{sec:one}--\ref{sec:three} are all for "mono-energetic" clusters, that is, clusters in which all the stars have a common semimajor axis. Real black-hole star clusters contain stars with a wide range of semimajor axes. In this case, since semimajor axis is conserved by resonant relaxation the stars in each small interval of semimajor axis can be regarded as a subsystem with a fixed number of particles, in thermal equilibrium with other semimajor axis intervals and sharing a common inverse temperature $\beta$. However, they cannot be regarded as distinct canonical ensembles because the different semimajor axes also interact through their common gravitational field. This problem is circumvented in the model described in \S\ref{sec:four} by scaling the number and mass of stars per unit semimajor axis so that all of these subsystems are identical. 
Our calculations ignore relativistic effects, which are important in the inner parts of black-hole star clusters, where $rM_\star(r)\lesssim r_\bullet M_\bullet$ so relativistic precession dominates over precession due to self-gravity. In this region the assumptions on which we have based our discussion break down, for two reasons. First, the relativistic correction to the Hamiltonian, $H_\mathrm{gr}=-(3/4)c^2 (r_\bullet/a)^2/(1-e^2)^{1/2}$, becomes large and negative for high-eccentricity orbits, so the distribution function $\sim\exp(-\beta H)$ diverges strongly if the inverse temperature is positive. Second, the relativistic precession rate diverges as $e\to 1$ so resonant relaxation becomes ineffective at high eccentricities (the Schwarzschild barrier). Thus there is a three-way tension between the relativistic Hamiltonian, which strongly favors radial orbits; the loss cone due to the black hole, which strongly disfavors them; and relativistic precession, which tends to freeze their evolution. Finally, we have restricted ourselves to clusters with zero total angular momentum. A similar transition in rotating axisymmetric clusters, if present, could explain the lopsided cluster at the center of the M31 galaxy \cite{kb99,kt13,bm13}. We have also not explored the possibility that the disordered and ordered states may coexist in a given cluster.

Black-hole star clusters are unresolved in optical/infrared images, except in the Milky Way and the nearest external galaxies. Thus lopsided clusters cannot generally be distinguished by their spatial structure. Instead, the primary observational consequence of the phase transition would be an increase in the rate of transient events such as flares from tidally disrupted stars and gravitational-wave signals from stars spiraling into the black hole. This increase arises because high-eccentricity orbits that bring stars close to the black hole are much more common in the lopsided state and is visible in Figure \ref{fig:two} as a rapid decline in the number of surviving stars once the mean eccentricity vector starts to grow. 

We have shown that black-hole star clusters undergo a robust phase transition from a disordered spherical state to an ordered, lopsided state when they are cooled below a critical dynamical temperature. In the disordered state, the rms eccentricity varies monotonically with temperature so the critical temperature can be expressed as an rms eccentricity, which in our calculations lies in the range 0.5--0.6. Several dynamical mechanisms could produce these low temperatures. These include removal of stars on high-eccentricity orbits, formation of stars on low-eccentricity orbits, the addition of stars on low-eccentricity orbits through the inspiral and tidal disruption of globular clusters, or diffusion of heat between stellar populations at different semimajor axes through resonant relaxation. Further investigation of cluster formation and evolution is needed to determine whether the phase transition occurs in realistic black-hole star clusters. 

\begin{acknowledgments}

We thank the anonymous referees for comments that sharpened our arguments and improved our presentation of them. 

J.T. acknowledges support from the Institute for Advanced Study through a Visiting Membership in the fall term of 2017. The simulations in this work were carried out on machines at the Dutch national e-infrastructure with the support of the SURF Cooperative through NWO grant 16109, as well as machines at the Institute for Advanced Study, the American University of Beirut, and Leiden Observatory. 

J.T. devised the project and was in charge of overall direction and planning. The calculations in \S\ref{sec:one} and \S\ref{sec:four} were conceived and carried out by S.T. The calculations in \S\ref{sec:two} and \S\ref{sec:three} were conceived by J.T., who further contributed to the construction of related numerical algorithms. M.K. implemented and optimized these algorithms, then performed the calculations and analyzed the results. J.T. and S.T. constructed the theoretical framework for the project. S.T. took the lead in writing the manuscript, with critical input from J.T.\ and M.K. 

\end{acknowledgments}


\begin{thebibliography}{1}

\bibitem{kh13} J.\ Kormendy and L.\ C.\ Ho, Coevolution (or not) of supermassive black holes and host galaxies. {Annu.\ Rev.\ Astron.\ Astrophys.}\ {\bf 51}, 511--653 (2013).

\bibitem{rt96} K.\ P.\ Rauch and S.\ Tremaine, Resonant relaxation in stellar systems. {New Astron.} {\bf 1}, 149--170 (1996).

\bibitem{st16} S.\ Sridhar and J.~R.\ Touma, Stellar dynamics around a massive black hole - II. Resonant relaxation. {Mon.\ Not.\ R.\ Astron.\ Soc.} {\bf 458}, 4143--4161 (2016).

\bibitem{bof18} B.\ Bar-Or and J.-B.\ Fouvry,  Scalar resonant relaxation of stars around a massive black hole. {Astrophys.\ J. Lett.} {\bf 860}, L23, 9 pp. (2018).

\bibitem{tal17} T.\ Alexander, Stellar dynamics and stellar phenomena near a massive black hole.  {Annu.\ Rev.\ Astron.\ Astrophys.}\ {\bf 55}, 17--57 (2017). 

\bibitem{kt11} B.\ Kocsis and S.\ Tremaine, Resonant relaxation and the warp of the stellar disc in the Galactic Center. {Mon.\ Not.\ R.\ Astron.\ Soc.}\ {\bf 412}, 187--207 (2011).

\bibitem{geg10} R.\ Genzel, F.\ Eisenhauer, and S.\ Gillessen, The Galactic Center massive black hole and nuclear star cluster. {Rev.\ Mod.\ Phys.}\ {\bf 82}, 3121--3195 (2010).

\bibitem{bw76} J.\ N.\ Bahcall and R.~A.\ Wolf, Star distribution around a massive black hole in a globular cluster. {Astrophys.\ J.}\ {\bf 209}, 214--232 (1976).

\bibitem{ck78} H.\ Cohn and R.~M.\ Kulsrud, The stellar distribution around a black hole -- Numerical integration of the Fokker-Planck equation. {Astrophys.\ J.}\ {\bf 226}, 1087--1108 (1978).

\bibitem{footnote0} The existence of well-defined maximum-entropy states is exceptional in self-gravitating systems, which can increase their entropy without limit by placing a small amount of mass on weakly bound or unbound orbits. This behavior is not possible on the resonant-relaxation timescale because semimajor axes are conserved. 

\bibitem{footnote2} The following argument may provide some intuition for the physics of the phase transition. Lopsided states are possible because eccentric orbits in a Kepler potential do not precess, so the self-gravity of the stars can corral the apsidal lines of the stellar orbits into an aligned configuration. The orbit-averaged gravitational energy between two orbits diverges to $-\infty$ if all their osculating elements agree, so low energies can be achieved either if the orbits are nearly circular but randomly oriented, or if they are highly eccentric with aligned apsides. The latter configuration has lower orientational entropy but this can be outweighed by the higher entropy associated with the distribution in eccentricity or angular momentum.

\bibitem{ttk09} J.~R.\ Touma, S.\ Tremaine, and M.~V.\ Kazandjian,  Gauss's method for secular dynamics, softened. {Mon.\ Not.\ R.\ Astron.\ Soc.}\ {\bf 394}, 1085--1108 (2009).

\bibitem{tt14} Touma, J., \& Tremaine, S., The statistical mechanics of self-gravitating Keplerian disks. {J.\ Phys. A}\ {\bf 47}, 292001 (2014)

\bibitem{cre83} M.\ Creutz, Microcanonical Monte Carlo simulation. {Phys.\ Rev.\ Lett.}\ {\bf 50}, 1411--1414 (1983).

\bibitem{tre05} S.\ Tremaine, Secular stability and instability in stellar systems surrounding massive objects. {Astrophys.\ J.} {\bf 625}, 143--155 (2005).

\bibitem{footnote1}Of course, a cluster could be formed with a non-monotonic phase-space distribution in angular momentum that is cold enough for dynamical instability. In such a case, lopsidedness may arise on the secular timescale, long before resonant relaxation leads to a maximum-entropy state. 

\bibitem{kb99} J.\ Kormendy and R.\ Bender, The double nucleus and central black hole of M31. {Astrophys.\ J.}\ {\bf 522}, 772--792 (1999). 

\bibitem{kt13} M.~V.\ Kazandjian and J.~R.\ Touma, The doubling of stellar black hole nuclei. {Mon.\ Not.\ R.\ Astron.\ Soc.}\ {\bf 430}, 2732--2738 (2013).

\bibitem{bm13} C.~K.\ Brown and J.\ Magorrian,  Three-dimensional Keplerian orbit-superposition models of the nucleus of M31. {Mon.\ Not.\ R.\ Astron.\ Soc.}\ {\bf 431}, 80--91 (2013). 

\end{thebibliography}
 \end{document}